\documentstyle[11pt,newpasp,twoside,epsf]{article}
\begin{document}

\title{On the Transition from AGB Stars to Planetaries: \\The Spherical Case}
 \author{Detlef Sch\"onberner and Matthias Steffen}
\affil{Astrophysikalisches Institut Potsdam, An der Sternwarte 16, \\
       D-14482 Potsdam}
%%\author{Matthias Steffen}
%%\affil{Astrophysikalisches Institut Potsdam, An der Sternwarte 16, 
%%       D-14482 Potsdam}

\begin{abstract}
    We discuss the basic physical model and the relevant 
    processes responsible for creating and shaping planetary nebulae
    out of a cool AGB wind envelope. We show that a  
    hydrodynamical treatment along the upper AGB leads quite naturally to more
    realistic starting configurations for planetaries with density slopes
    steeper than  $ r^{-2} $. Taking into account photoionization and wind
    interaction in a realistic manner, the hydrodynamics of post-AGB wind
    envelopes leads to density structures and velocity fields in close 
    resemblance to observations of spherical or elliptical planetary nebulae.
\end{abstract}

\section{Introduction}     \label{intro}

%%%\subsection{Why Spherical Studies?}    \label{why}
   Although we are here only interested in an explanation of non-spherical
   structures observed so often in planetary nebulae (PN), a detailed study
   of spherical systems appears to be important for at least two reasons. The
   first is a more physical one and refers to our still rather poor knowledge 
   of PN formation and evolution. The use of spherical models allows a detailed
   study of basic physical processes without having to worry about influences 
   caused by non-spherical structures. The other reason is a technical one: the
   presently available computing power is too limited to follow the
   evolution of non-spherical model planetaries over their whole life 
   with sophisticated physics and good spatial resolution.
   
   It is expected that basic physical processes work similarly in systems with 
   a complex geometry. They set the stage for the other phenomena
   responsible for the development of non-spherical structures and should
   always be considered.

\section{The Basic Physical System}  \label{basic}
\begin{figure}[h]                    %%%%%%%%%%%%%  Figure 1
  \plotfiddle{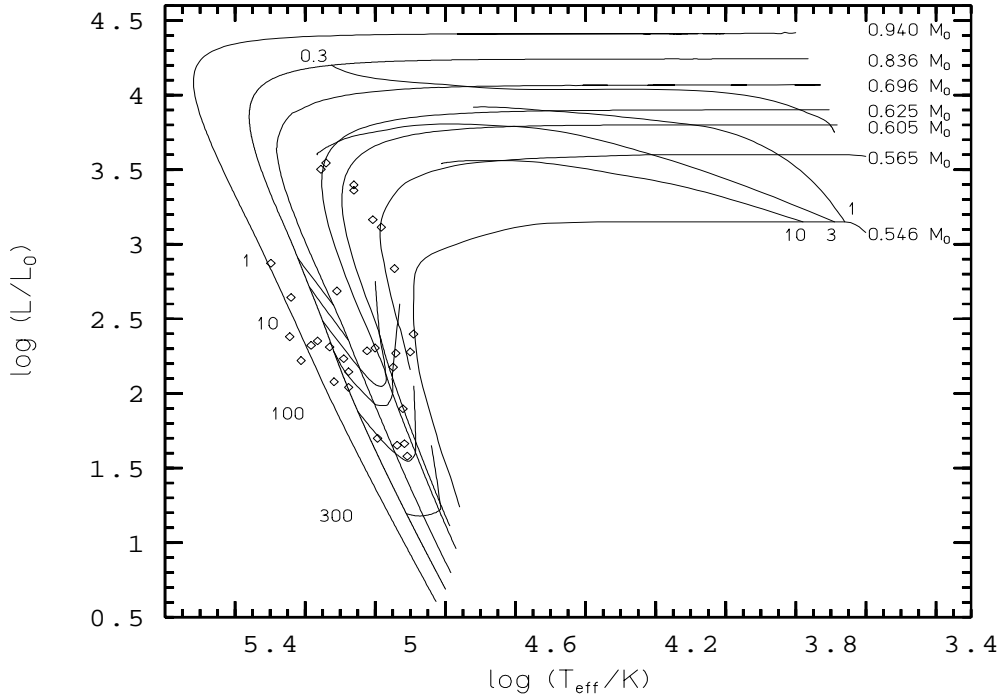}{9cm}{-90}{57}{52}{-220}{285}
  \caption{
           Evolutionary tracks of hydrogen-burning post-AGB models with 
           isochrones (in $10^3$~yrs), together  
           with a sample of central stars of `optically thick' PNe, from 
           Bl\"ocker (1995). The post-AGB ages are counted from positions close
           to the AGB and may depend on the assumed mass-loss rates in
           the vicinity of the AGB (cf.\ Bl\"ocker 1995).
          }     \label{tracks}
\end{figure}

   The evolution of an AGB star is driven by mass loss until the mantle
   is lost and the remnant begins to contract rapidly towards higher 
   temperatures. Eventually the burning shells extinguish and the white dwarf 
   cooling path is reached. The remnant's luminosity and evolutionary speed
   depend very sensitively on its mass, and possible ranges of luminosity
   and speed are shown in Fig.~\ref{tracks}. All these remnants stem from
   different progenitors whose evolutionary histories have been consistently
   followed from the main sequence through all the later phases including mass 
   loss and thermal pulses (see Bl\"ocker 1995 for the details). Similar
   computations have been performed by Vassiliades \& Wood (1994). 
%%%%%%%%%%%%%%%%%%%%%%%%%%%%%%%%%%%%%%%%%%%%%%%%%%%%%%%%%%%%%%%%%%%%%%%%%%%%%%%
%%%%%%%%%%%%%%%%%%%%%%%%%%%%%%%%%%%%%%%%%%%%%%%%%%%%%%%%%%%%%%%%%%%%%%%%%%%%%%%   
  
   The evolution of the AGB remnant, - the
   central star -, in temperature and luminosity drives in turn the 
   development of a PN out of a cool wind envelope
   by {\em two\/} processes, viz.\ by the concomitant changes of the stellar 
   radiation field and wind power.  The relative importance
   of both processes with respect to the evolution of a planetary   
   varies with the central-star's age (or effective temperature).   
    The number of hydrogen-ionizing photons emitted per second increases  
   rapidly with the 
   remnant's effective temperature, but later the luminosity decrease starts to
   dominate. For a typical central-star mass of 0.6~M$_{\sun}$ the maximum
   flux of ionizing photons occurs between 60\,000 and 70\,000~K. The very
   rapid luminosity drop after the central star has reached its maximum
   effective temperature (cf.\ Fig.~\ref{tracks}) may cause substantial
   recombination. It is important to follow this late evolutionary phase with
   a fully time-dependent code that treats all the relevant physical processes, 
   i.e.\ ionization, recombination, heating and cooling (Marten 1995). 
   
   The property of mass-loss during the post-AGB evolution is more
   difficult to evaluate. 
   For an AGB star we have winds driven by radiation pressure on small grains
  with momentum transfer to the gas. The outflow rates depend on the star's
  luminosity and effective temperature (cf.\ Sedlmayr \& Dominik 1995; Arndt,
  Fleischer, \& Sedlmayr 1998). Typical rates are between about $ 10^{-7} $ and 
  $ 10^{-4}$~M$_{\odot}$, with outflow velocities from 5 to 25~km/s, i.e.\ 
  $\la V_{\rm esc}$, the surface escape velocity. During
  the post-AGB contraction, mass-loss rates are lower by orders of magnitude,
  but the wind velocities are substantially higher. The driving of the outflow
  occurs via radiation pressure on lines (cf.\ Pauldrach et al.\ 1988), 
  $\dot{M} \simeq 1.3 \cdot 10^{-15}\,(L/L_{\odot})^{1.86}$,
   and typical values are $ \approx 10^{-8}$~M$_{\odot}$/yr for the rate, but 
   now with $ V \simeq 1\,000 \dots 10\,000$ km/s $ \simeq (2 \dots 4) 
   V_{\rm esc}$. The wind power, $ P = \dot{M} V^2 /2 $, reaches its
   maximum close to the turn-around point at maximum effective temperature and
   declines then rapidly with the luminosity.
 
   It should, however, be noted that the stellar wind does not interact 
   directly with the nebular/AGB material. Instead, the wind's kinetic energy 
   thermalizes 
   through a shock and adds to the energy and matter content of hot, shocked 
   wind material emitted at earlier times.  The thermal pressure of this 
   `bubble' of hot but very tenuous gas drives the inner edge of the 
   planetary. Though it is actually the time integral over the wind power that
   determines the energy content of the bubble, the maximum bubble pressure
   coincides roughly with the maximum wind power of the central star.

\section{Formation and Evolution of PN}   \label{formevol}

   Given typical mass-loss rates between $ 10^{-5} $ and 
   $ 10^{-4}$~M$_{\sun}$/yr during the final AGB evolution and the still rather
   low wind velocities during the remnant's transition through the cool part
   of the Hertzsprung Russell diagram, the dynamical effects of wind 
   interaction are expected to be modest. As soon as the remnant 
   becomes sufficiently
   hot, ionization sets in and gives birth to a HII region deeply embedded in 
   the neutral/molecular circumstellar AGB material. Thermal pressure of the
   ionized  matter drives a shock wave into the ambient slow material, and the
   front itself defines the outer edge, $ R_{\rm pn}$, of the new PN
   even if the
   ionization has already broken through into the surrounding region. The 
   front's speed, $ \dot{R}_{\rm pn}$, 
   is mainly determined by the balance of the shell's thermal pressure 
  %%% $P_{\rm shell}$, 
   with  the ram pressure exerted by the ambient matter.
   At a given time, speed and position of the outer rim of a planetary depend 
   thus on the mass-loss history over the last 10\,000 to 20\,000~years of 
   AGB evolution. The mass embraced by $ R_{\rm pn}$ is steadily growing
   with time at the expense of the still undisturbed (although possibly ionized) 
   AGB wind material. 
   We emphasize here that $ \dot{R}_{\rm pn}$ is {\em not\/} a matter 
   velocity and cannot be observed spectroscopically! 
   
   As outlined in Sect.~\ref{basic} above, the wind interaction through the
   hot bubble becomes more and more important with time and compresses and 
   accelerates the inner parts of the shell into a high-density shell, the
   so-called `rim' (cf.\ Balick 1987). Since the bubble's pressure is 
   controlled by the central-star's wind properties, the evolution of the 
   central star controls the shaping of the inner parts of a planetary.
   
   Hydrodynamical simulations that took ionization and wind 
   interaction properly into account have shown that both effects 
   lead unavoidably to typical double-shell structures consisting of an inner 
   high-density `rim' surrounded by a low-density `shell' 
    with no resemblance to the initial density and velocity 
   distributions (cf.\ Marten \& Sch\"onberner 1991; Mellema 1994). 
   Thus planetaries {\em do not\/} contain direct information on precedings
   mass-loss phases during the end of the AGB evolution!

\section{Two-Component Radiation Hydrodynamics Simulations of the \\Final 
         AGB Phase}                                                \label{sim}

   Attempts to model the evolution of planetary nebulae face the problem of
   selecting the proper initial configuration, i.e.\ density distribution and 
   velocity field. Since practically nothing is known, rather simple
   conditions are usually assumed, viz.\ mass outflow with constant speed and 
   rate. Our present knowledge of the late stages of stellar evolution allows, 
   however, to draw more detailed conclusions: 
   (i) The theory of radiation-driven winds on the AGB suggests that 
       both the outflow rate and -speed depend on the stellar luminosity and 
       effective temperature, and on the chemical composition as well (Arndt 
       et al.\ 1997).  
   (ii) Stellar evolution theory predicts large luminosity  variations 
        (up to a factor three) during thermal pulses, expected to lead to 
        drastic variations of outflow rates and speeds. 
        
   One can expect that hydrodynamical simulations of AGB wind envelopes along 
   the upper AGB give very 
   useful informations about initial conditions to be expected for planetaries.  
   A first step into this direction has been reported by Sch\"onberner et al.\
   (1997). The stellar outflow is assumed to be spherically symmetric,
   and the equations of hydrodynamics are solved for the gas and the dust
   component, coupled by momentum exchange due to dust-gas collisions.  We used 
   a modified version of the code developed by Yorke \& Kr\"ugel (1977),  
   making use of the following simplifications: 
   (i) Radiation transfer is considered only for the dust component, i.e.\
       exchange of photons between dust grains and the gas is neglected.  
   (ii) The dust temperature is computed from radiative equilibrium, and
        the gas  (neutral hydrogen) is assumed to have the same
        (local) temperature.  
   (iii) The dust consists of single-sized grains, either based on
         oxygen or carbon chemistry, adopting a fixed dust-to-gas ratio
         at the dust condensation point.
 
   We introduced time-dependent values of stellar mass,
   luminosity, effective temperature {\em and\/}  variable mass loss
   (as shown in Fig.~\ref{Off}) 
   with a constant flow velocity equaling the local sound speed, $\approx
   3$~km/s as a boundary condition. The radiation pressure on the grains and the
   momentum transfer to
   the gas leads to an acceleration of the material to typical final outflow
   velocities around 10 to 15 km/s, in agreement with observations.
   A more detailed description of this fully implicit radiation hydrodynamics 
   code has been given by Steffen et al.\ (1997) and Steffen, Szczerba, 
   \& Sch\"onberner (1998).
      
\subsection{Evolution through the Upper AGB and Beyond}   \label{beyond}  
  
\begin{figure}[h]                %%%%%% Figure 2  
     \plotfiddle{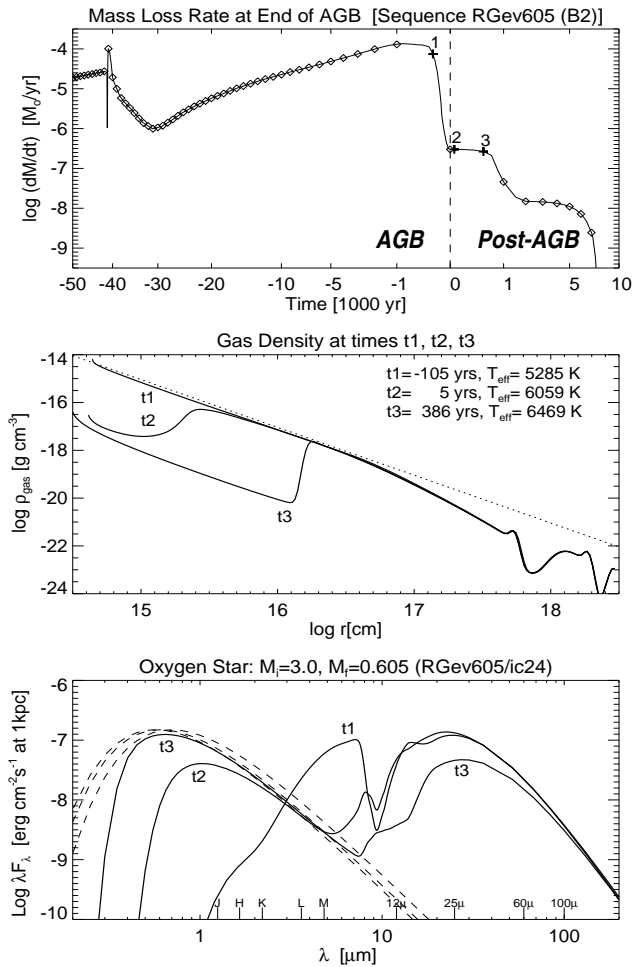}{12cm}{0}{55}{52}{-160}{-32}
     \caption{ \label{Off}
      {\bf Top:} Adopted mass-loss rate for the late AGB and early post-AGB
      evolution ending up with a remnant of 0.605 M$_{\odot}$ (Bl\"ocker 1995).
      {\bf Middle:} Radial gas density distributions for models with silicate
      grains at the three selected times marked in the upper panel.
      {\bf Bottom:} Spectral energy distributions at the same times. The dashed 
      lines are the corresponding intrinsic spectra of the central star.       
           }
   \end{figure}

 We extended our AGB hydrodynamical simulations somewhat into the post-AGB 
   regime,
   using the mass-loss prescription shown in the upper panel of Fig.~\ref{Off}.
   Mass-loss rate and effective temperature (or radius) of the star are coupled 
   according to
   the prescription of Bl\"ocker (1995), and the most prominent feature is a 
   rapid decrease of the rate by orders of magnitude within about 100 years 
   around effective temperatures of 6\,000~K. The consequence is
   a rapid detachment and thinning of the dust shell since the density of any
   newly formed dust is strongly reduced and gives no detectable signature. 
   This is illustrated  by the sequence of spectral energy distributions
in the lower panel of Fig.~\ref{Off} which covers a time interval of less than
   500 years. For this simulation we adopted an oxygen-based grain type
   (``Astronomical Silicates''), and the gradual disappearance of the strong
silicate absorption feature with increasing shell detachment is clearly seen. At
   the same time, the previously totally obscured AGB remnant becomes visible. 
   Our modelled spectral energy distributions resemble very much those of
   known proto-planetary nebulae (Hrivnak, Kwok, \& Volk 1989), indicating
   that the mass-loss variations at the end of the AGB evolution as chosen by 
   Bl\"ocker (1995) are close to reality!
%%%%%%%%%%%%%%%%%%%%%%%%%%%%%%%%%%%%%%%%%%%%%%%%%%%%%%%%%%%%%%%%%%%%%%%%%%%%%%%
%%%%%%%%%%%%%%%%%%%%%%%%%%%%%%%%%%%%%%%%%%%%%%%%%%%%%%%%%%%%%%%%%%%%%%%%%%%%%%

   Due to the variations of the mass-loss rate as shown in the upper panel of
   Fig.~\ref{Off}, the density structure is clearly
   different from the usual assumption of a $\rho \propto r^{-2}$ law 
   (middle panel): The  density dip near $r = 10^{18}$~cm 
   is caused by the last thermal pulse about 30\,000 years ago 
   (cf.\ upper panel), while the rapid density increase towards the inner 
   parts of
   the shell ($\rho \propto r^{-3}$) is due to the recent increase of mass-loss
   rate. Further inwards the density increase flattens somewhat ($\rho \propto
   r^{-1}$). The outflow velocity is rather constant, $\approx 11$~km/s,
   except for a slight decrease during the last thermal pulse.

\subsection{Evolution across the Hertzsprung Russell Diagram}     \label{HRD}  
  
 Little is really known about the development of wind strength and speed 
   during the
   early post-AGB evolution. In the model shown in Fig.~\ref{Off} the mass loss 
   is set to the Reimers prescription (Reimers 1977)
   which is then kept until the remnant  becomes hot enough for
   the theory of radiation driven winds to be applicable 
   (Pauldrach et al.\ 1988). A more detailed description  
   how mass-loss rate and wind speed may vary in the course of the post-AGB
   evolution is given in Marten \& Sch\"onberner (1991).

   In order to investigate the transformation of a cool AGB wind envelope into
   a planetary nebula, we used the model structure shown in Fig.~\ref{Off}
   at time $ t2 $ as input for another radiation hydrodynamics code. This 
   one-component explicit code is 
   based on a second-order Godunov-type advection scheme and 
   considers time-dependent ionization, recombination, heating and
   cooling of six elements (H, He, C, N, O, Ne) with all of their ionization
   stages. More details can be found in Perinotto et al.\ (1998).
%%%%%%%%%%%%%%%%%%%%%%%%%%%%%%%%%%%%%%%%%%%%%%%%%%%%%%%%%%%%%%%%%%%%%%%%%%%%%%%
   \begin{figure}[hb]                %%%%%% Figure 3  
      \plotfiddle{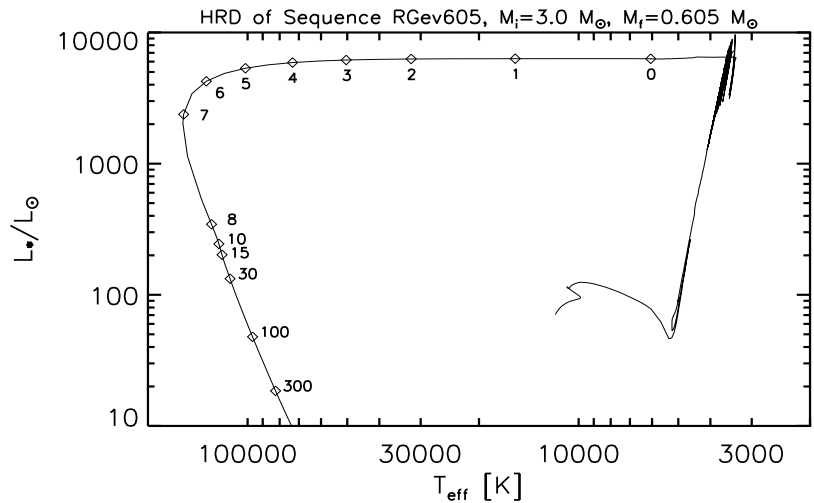}{6cm}{0}{60}{57}{-180}{-43}
      \caption{
            Complete evolutionary track of the 3~M$_{\sun}$ model ending up
            as a 0.605~M$_{\sun}$ white dwarf as used in our numerical 
            simulations (from Bl\"ocker 1995). Time marks (in $10^3$~yrs) 
            along the post-AGB path correspond to the post-AGB time scale 
            given in Fig.~\ref{Off}. 
              }      \label{605}
   \end{figure}
%%%%%%%%%%%%%%%%%%%%%%%%%%%%%%%%%%%%%%%%%%%%%%%%%%%%%%%%%%%%%%%%%%%%%%%%%%%%%%
%%%%%%%%%%%%%%%%%%%%%%%%%%%%%%%%%%%%%%%%%%%%%%%%%%%%%%%%%%%%%%%%%%%%%%%%%%%%%%
   \begin{figure}[h]                %%%%%% Figure 4 
      \plotfiddle{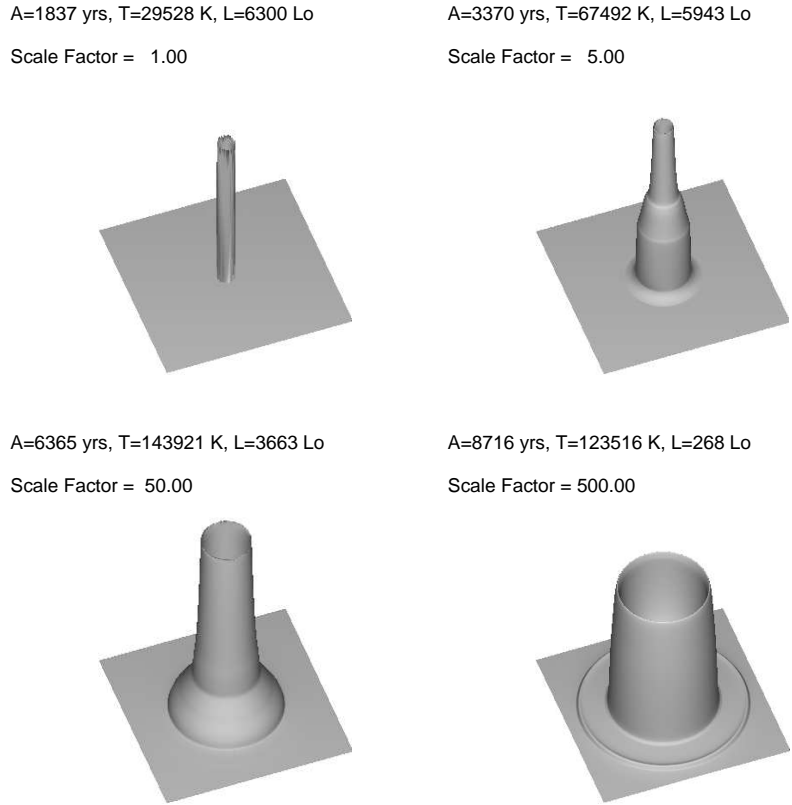}{11cm}{0}{60}{60}{-170}{-30}
      \caption{
           3D representation of the surface-brightness distributions in 
           H$\alpha$\ of selected models 
           along the post-AGB track displayed in Fig.~\ref{605}. The models
           are labelled by their post-AGB age, effective temperature and
           luminosity of their central stars. Scale factors were applied
           in order to compensate for the decrease of the surface 
           brightness with time.
              }      \label{surface}    
   \end{figure}
%%%%%%%%%%%%%%%%%%%%%%%%%%%%%%%%%%%%%%%%%%%%%%%%%%%%%%%%%%%%%%%%%%%%%%%%%%%%%%%

   A visualization how our model planetary develops in size, brightness and
   structure is presented in Fig.~\ref{surface}, showing H$\alpha$ 
   surface-brightness distributions for selected models taken from our 
   hydrodynamical simulation along the post-AGB evolutionary path displayed
   in Fig.~\ref{605}. 
   
   At age = 1\,837~yrs (upper left in Fig.~\ref{surface}) 
   ionization has already created a small but bright shell limited by a density
   wave which keeps the photons  trapped. The peak flow velocity in this wave
   is 23~km/s, whereas the flow at the inner edge of the shell is nearly
   stalling with only about 4~km/s.
   About 1\,500 years later (upper right) the ionization has broken through the 
   shock, and the shell expands and dilutes because 
   the wave front (the shock) is further accelerated due to the steeper than
   $ \rho^{-2}$ density slope. 
   At about this time, wind interaction becomes noticeable through the 
   formation of a compressed, bright `rim' at the inner edge of the shell. 
   
   At age = 6\,365~yrs (lower left), the brightness contrast between inner
   rim and shell has
   significantly increased by the combined action of the shell's expansion into 
   the AGB wind and compression from inside by the `hot bubble'. The whole
   structure corresponds to a typical `attached-halo multiple-shell PN' 
   (Stanghellini \& Pasquali 1995). The maximum flow velocity,  
   immediately behind the shock front,  
   is now 32~km/s, that of the rim matter about 24~km/s. This model
   agrees also qualitatively with the results of a structural and kinematical
   study of G\c esicki, Acker, \& Szczerba (1996) who found, e.g., for the
   double-shell planetary IC~3568 shell velocities up to 40~km/s, but only
   about 10~km/s for the inner dense parts (the rim).
      
   When the central star's luminosity has dropped rapidly to only a few 
   100~L$_{\sun}$, recombination within the shell reduces its brightness to 
   typical  halo values (age = 8716~yrs, lower right) which ends then the double
   shell phase that lasted from about age = 2800 till age = 7600~yrs, i.e.\ for
   a quite substantial fraction of a typical PN life time. Though the shell's
   brightness compares now with that of a halo, it is not a halo: the matter 
   within the recombined shell continues to expand and compresses the AGB gas
   into a dense but thin shell, leading to substantial limb brightening. An
   example for such a structure is NGC~2438 which consists of a bright 
   ring-like shell surrounded by a limb-brightened `halo'. The analysis of
   Corradi et al.\ (in preparation) shows that this `halo' is actually the
   recombined former shell set up by ionization at the very beginning of the
   planetary's life.

\end{document}